\documentclass[12pt]{article}
\usepackage{setspace}
\usepackage[T1]{fontenc}
\usepackage[utf8]{inputenc}
\usepackage{authblk}

\usepackage{amsmath,amsfonts,amssymb,enumerate}
\usepackage{setspace}
\usepackage{color}
\usepackage{lscape}
\usepackage{float}
\usepackage{booktabs}
\usepackage{multirow}
\usepackage{pifont}
\usepackage{diagbox}
\usepackage{tabularx}
\usepackage{extarrows}
\usepackage{mathrsfs}
\usepackage{indentfirst} 
\usepackage{autonum}
\usepackage{rotating}
\usepackage{threeparttable}
\usepackage{graphicx}
\usepackage{epstopdf}
\usepackage{caption}
\usepackage{epsfig}
\usepackage{subfigure}
\usepackage{algpseudocode,algorithm}
\usepackage[margin=1in]{geometry}
\usepackage{verbatim}
\usepackage{bm}
\usepackage{bbm}
\usepackage{cases}
\usepackage{indentfirst}
\newtheorem{theorem}{\bf Theorem}[section]
\newtheorem{lemma}[theorem]{\bf Lemma}
\newtheorem{proposition}[theorem]{\bf Proposition}

\newenvironment{proof}{\noindent{\em Proof:}}{\quad \hfill$\Box$\vspace{2ex}}

\title{Gaussian graphical models with graph constraints for magnetic moment interaction in high entropy alloys}
]
\author[1]{Xinrui Liu}
\author[2]{Yifeng Wu}
\author[2]{Douglas L. Irving}
\author[3]{Meng Li}
\affil[1]{School of Mathematics and Statistics, Shandong Normal University, Jinan, 250358, China}
\affil[2]{Department of Materials Science and Engineering, North Carolina State University, 911 Partners Way, Raleigh, NC, 27606, USA.}
\affil[3]{Department of Statistics, Rice University, 6100 Main Street, MS 138, Houston, TX 77005, USA.}

\date{}

\usepackage{natbib}
\begin{document}
\maketitle
\vspace*{-1cm}
\begin{abstract}
This article is motivated by studying the interaction of magnetic moments in high entropy alloys (HEAs), which plays an important role in guiding HEA designs in materials science. While first principles simulations can capture magnetic moments of individual atoms, explicit models are required to analyze their interactions. This is essentially an inverse covariance matrix estimation problem. Unlike most of the literature on graphical models, the inverse covariance matrix possesses inherent structural constraints encompassing node types, topological distance of nodes, and partially specified conditional dependence patterns. The present article is, to our knowledge, the first to consider such intricate structures in graphical models. In particular, we introduce graph constraints to formulate these structures that arise from domain knowledge and are critical for interpretability, which leads to a Bayesian conditional autoregressive model with graph constraints (CARGO) for structured inverse covariance matrices. 
The CARGO method enjoys efficient implementation with a modified Gauss-Seidel scheme through proximity operators for closed-form posterior exploration. 
We establish algorithmic convergence for the proposed algorithm under a verifiable stopping criterion. Simulations show competitive performance of CARGO relative to several other methods and confirm our convergence analysis.
In a novel real data application to HEAs, the proposed methods lead to data-driven quantification and interpretation of magnetic moment interactions with high tractability and transferability. 

\end{abstract}
\section{Introduction}
High entropy alloys, or HEAs~\citep{cantor2004microstructural, yeh2004nanostructured}, have received considerable attention in materials science and engineering. Unlike conventional alloys which consist of a principal element with minor additions, HEAs consist of five or more principal elements in equimolar or near-equimolar ratios, making it chemically complex at crystalline lattice sites. This also stimulates explorations of lower-order systems including ternary and quaternary alloys, which are now, together with HEAs, broadly referred to as complex concentrated alloys (CCAs)~\citep{Miracle2017,George2019high} or multiple principle element alloys (MPEAs). HEAs with such chemical complexity exhibit desired performance for mechanical properties \citep{Zhang2014,2019Mechanical}, corrosion resistance \citep{Lu2018, 2019Corrosion}, etc. The NiCrCoFe medium-entropy alloy (MEA) serves as a critical subsystem among the 3D transition metal HEA family.
This MEA itself has shown intriguing magnetostructural~\citep{Niu2015} and magnetomechanical~\citep{Niu2018Magnetically} behaviors, owing to rich interplays between the chemical chaos and atomic spins. These atomic spins are highly dependent on the local chemical environment and, consequently, sophisticated inter-atomic correlations. While first principles simulations can capture magnetic moments of individual atoms, explicit models are required to analyze their interactions. Therefore, one key question in the emerging field of materials informatics is whether or not a model-based approach, parsimonious and interpretable, can detect magnetic interactions in a data-driven fashion, so as to provide insights to the \emph{magnetism}-property-processing relationships in HEA designs.

Such a problem is closely related to spatial point pattern analysis in three-dimensional space. Spatial point pattern analysis has wide applications in a variety of areas, including geography, ecology, biology, and epidemiology \citep{gatrell1996spatial,Wiegand2013Handbook,Ghamisi2014Spectral,2014Spatial,li2019bayesian}. Analyzing magnetic interactions in HEA studies points to seeking the spatial correlation of magnetic moments between atoms in a proper spatial point pattern. In particular, we describe the atomic structure of HEAs by a Gaussian Markov random field (GMRF) \citep{balram1993noncausal,sorbye2014scaling} through a Gaussian conditional autoregressive (CAR) model \citep{besag1974spatial,besag1995on, lee2011a} for the spatial correlation of the central atom and its neighbors. We view each atom in HEAs as a node in a GMRF, and any two atoms (nodes) whose magnetic moments interact with each other are connected by an edge. Thus, characterizing the spatial correlation of the magnetic moments between atoms in HEAs boils down to estimating the inverse covariance matrix of a GMRF. 

There has been a rich literature on sparse inverse covariance matrix estimation in recent years \citep{NIPS2012_4574,NIPS2012_4601,cai2016estimating}. However, the entailed domain knowledge in HEAs poses unique challenges. For example, the position of the non-zero elements of the inverse covariance matrix can be completely determined by the atomic structure of a HEA. In addition, the type of each node (such as Ni, Cr, Co, or Fe) has a profound effect in describing the spatial correlation, with an added feature that a neighborhood in HEAs is defined in a topological sense owing to the periodic pattern of HEAs. We formulate these inherent structures that arise from domain knowledge into graph constraints. To our best knowledge, inference on inverse covariance matrices that complies with such graph constraints has not been addressed. 

Bayesian methods allow the incorporation of constraints in prior distributions while providing uncertainty quantification. This appealing strategy is often hampered by challenging posterior computation under complex constraints. To broaden the practicability of Bayesian modeling for HEAs, an efficient algorithm that enables fast computation for key posterior quantities is needed. 

In this article, we propose a Bayesian method, conditional autoregressive model with graph constraints (CARGO), to learn structured inverse covariance matrices. Using GMRFs to model the atomic magnetic structure of HEAs, we account for node types and topological structures to define node neighborhoods, and incorporate domain knowledge as graph constraints into the model. We employ a constrained Wishart prior distribution for the covariance structure. As far as we know, this is the first model-based approach for studying atomic interactions in HEAs. Under the constrained Wishart prior, the maximum a posterior (MAP) estimate in CARGO is turned to a optimization problem with graph constraints. 
We develop a penalty decomposition proximal modification of the Gauss-Seidel algorithm to carry out the optimization. We characterize the Gauss-Seidel iterative scheme using proximity operators, which leads to a deterministic algorithm with closed-form solutions, enabling fast computation. We establish a range of algorithmic convergence properties for the CARGO procedure. In particular, we show that the sequence generated by the proposed algorithm converges to the global minimum of the constrained optimization problem induced by MAP estimation under a verifiable stopping criterion.    

It is worth mentioning that \cite{2020Constrained} proposes an alternative strategy for constrained Bayesian inference by projecting unconstrained posterior distributions to the constrained parameter space. They show that this projection approach can be interpreted by data-dependent priors for certain likelihoods, while circumventing posterior computation under constraints that is often challenging. We instead use constrained Wishart prior distributions, which are independent of data, and employ tools in the optimization literature to establish fast, deterministic algorithms for posterior summary, reassured by extensive convergence analysis. 

In a novel real data application to detect atomic magnetic moment interactions, our model-based estimation agrees with the conclusions in the existing literature from materials science, leading to interpretable insights into HEA designs with high tractability and transferability. 

The rest of this article is organized as follows. We describe HEA data in Section \ref{sec:background.HEA}. Section \ref{sec:method} contains the Bayesian model of HEAs and the proposed constrained Wishart priors. We include a novel algorithm for posterior computation and carry out the convergence analysis in Section~\ref{sec:algorithm.convergence}. Section \ref{sec:simulation} assesses finite sample performance of the proposed method relative to other methods using simulations. We apply the proposed method to analyzing HEA data in Section \ref{sec:data.application}. We conclude this article in Section~\ref{sec:discuss}.

 \section{HEA data} 
\label{sec:background.HEA}

\begin{figure}[htbp]
\centering

\subfigure[fcc-1]{
\begin{minipage}[t]{0.15\linewidth}
\centering
\includegraphics[width=1in]{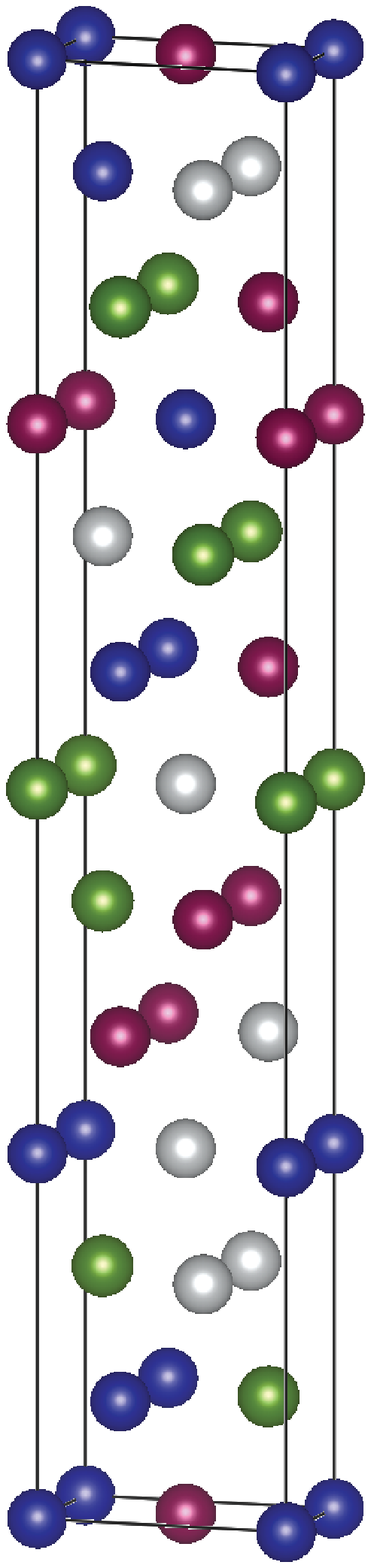}
\end{minipage}%
}%
\subfigure[fcc-2.]{
\begin{minipage}[t]{0.15\linewidth}
\centering
\includegraphics[width=1.01in]{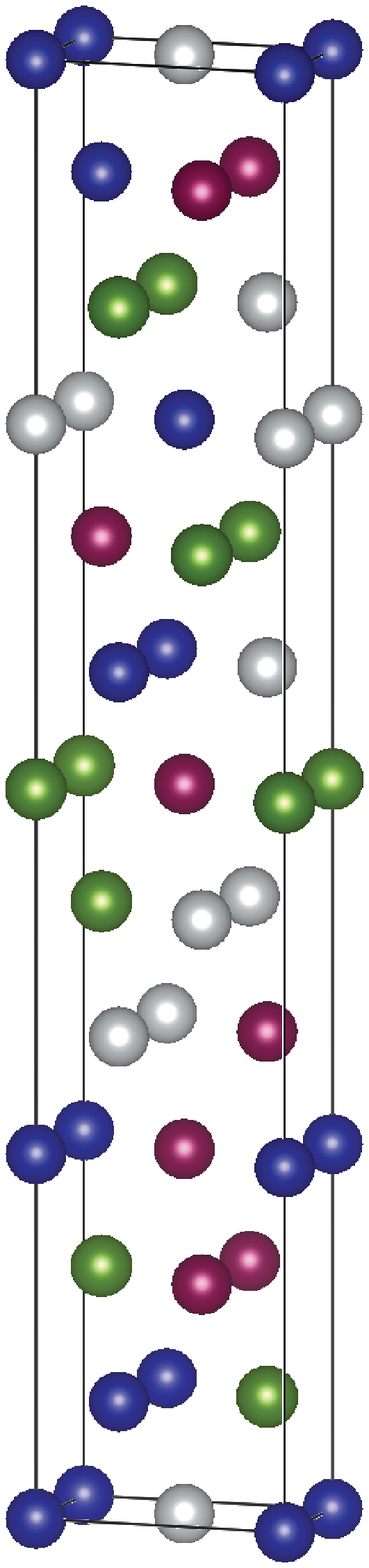}
\end{minipage}
}%
\subfigure[hcp-1.]{
\begin{minipage}[t]{0.15\linewidth}
\centering
\includegraphics[width=0.96in]{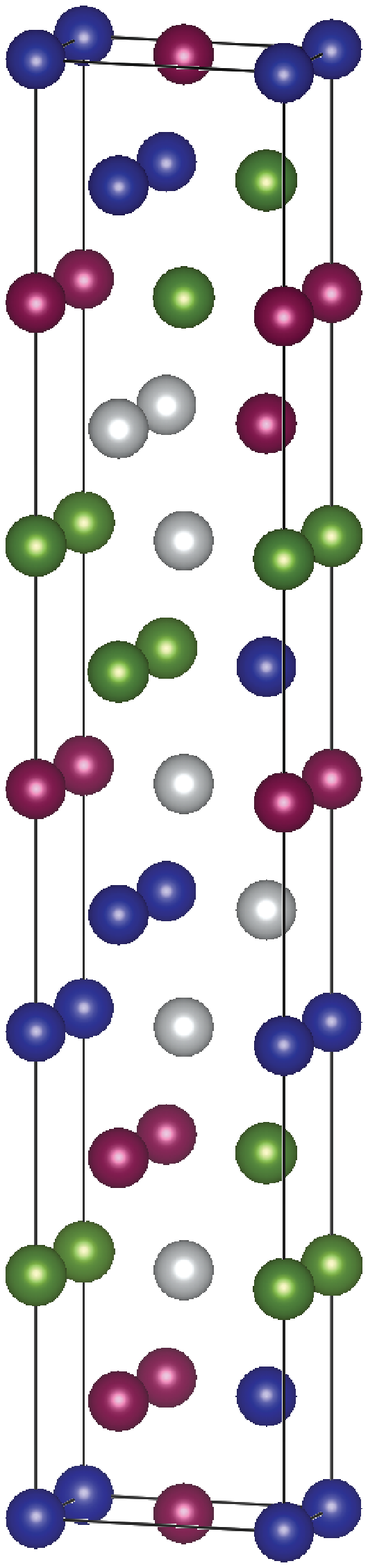}
\end{minipage}%
}%
\subfigure[hcp-2.]{
\begin{minipage}[t]{0.15\linewidth}
\centering
\includegraphics[width=0.96in]{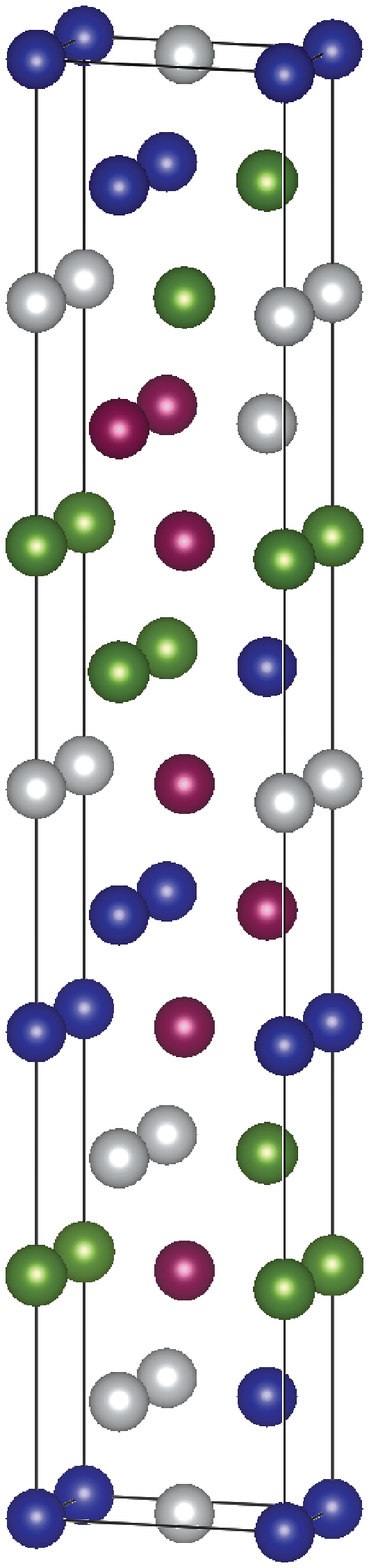}
\end{minipage}
}%
\subfigure[dhcp-1]{
\begin{minipage}[t]{0.15\linewidth}
\centering
\includegraphics[width=0.97in]{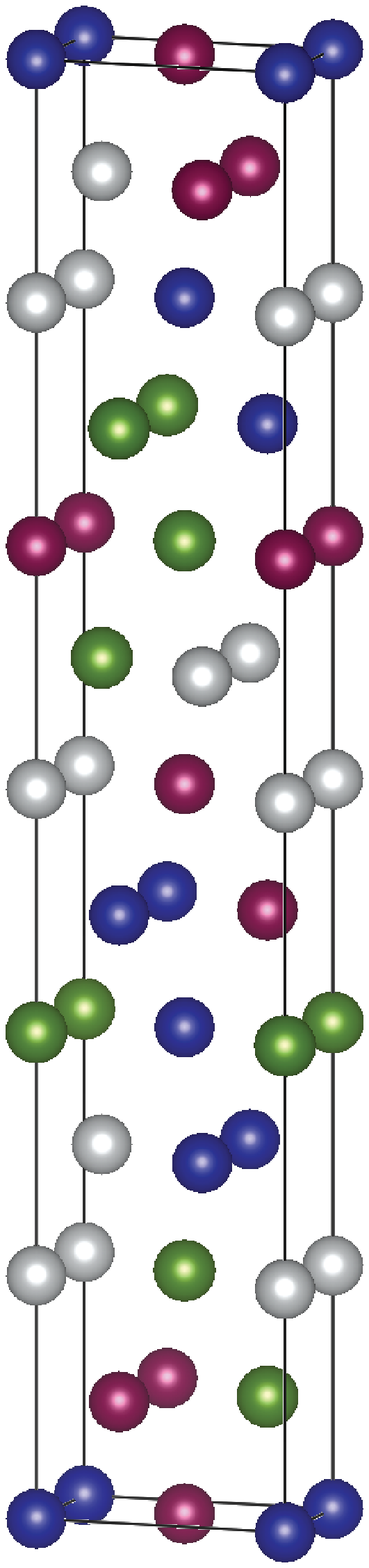}
\end{minipage}%
}%
\subfigure[dhcp-2.]{
\begin{minipage}[t]{0.15\linewidth}
\centering
\includegraphics[width=0.96in]{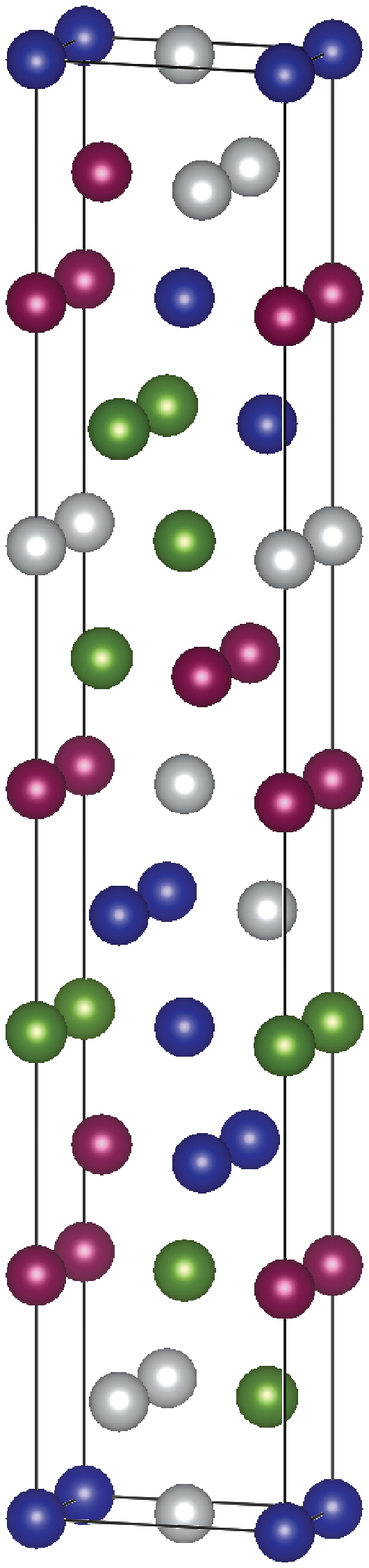}
\end{minipage}%
}%

\centering
\caption{Visualization of HEA supercells. We select two samples for each of the three HEA structures (fcc, hcp, and dhcp). Atoms in blue, green, red, and grey are Co, Cr, Fe, and Ni, respectively.}
\label{999999}
\end{figure}

Our real data application involves three different HEA structures: face centered cubic (fcc), hexagonal close packed (hcp), and double hcp (dhcp). Using the alloy theoretic automated toolkit (ATAT) \citep{van2002alloy}, a special quasi-random structure (SQS) \citep{zunger1990special} was generated for each of these. Each SQS or sample contains a 24-dimensional magnetic moment vector and an atom type vector. The four types of atoms are Ni, Co, Cr and Fe, short for Nickel, Cobalt, Chromium and Iron, respectively; atom types are distributed evenly, i.e., each sample consists of 6 Ni/Co/Cr/Fe atoms. This corresponds to an equiatomic NiCrCoFe system. Figures~\ref{999999} shows the 3D atomic structure of two samples for each of the three HEA structures.
Before introducing the geometric details, it should be noted that each supercell is a repeating unit of the complete fcc, hcp, or dhcp structure, such that the crystallographic periodicity needs to be considered in calculating its number of atoms, that is, atoms on faces, sides, and corners are counted as 1/2, 1/4, and 1/8, respectively.
These supercells were chosen to be orthorhombic, meaning that the axes \emph{x}, \emph{y}, and \emph{z} are perpendicular to each other.
Specifically, in each supercell, the \emph{x}-\emph{y} plane forms a 2-atom rectangle cutoff from the close-packed plane, with a stack of 12 close-packed planes rendered along the \emph{z}-axis.
Such a way of cutting through the crystal leads to a geometric condition of \emph{x}:\emph{y}:\emph{z}=1:$\sqrt{3}$:$4\sqrt{6}$.
Throughout these supercells, the in-plane arrangements of atoms involve 3 types, shortly denoted by A, B, and C.
The geometrical differences between fcc, hcp, and dhcp structures can thus be recognized by the sequence of atomic configurations along the \emph{z}-axis, which is $\cdots$ABCABC$\cdots$ in fcc, $\cdots$ABAB$\cdots$ in hcp, and $\cdots$ABACABAC$\cdots$ in dhcp.

Density functional theory (DFT) calculations, implemented in the Vienna Ab initio Simulation Package (VASP) \citep{kresse1993ab,kresse1994ab,kresse1996efficiency,kresse1996efficient}, were performed to obtain the ground-state atomic magnetic moments in these SQSs.
The generalized-gradient approximation (GGA) with Perdew-Burke-Ernzerhof (PBE) parametrization \citep{perdew1996generalized} for exchange-correlation functional was employed.
A 9$\times$5$\times$1 $\Gamma$-centered k-point mesh with a cutoff energy of 350 eV for the plane wave basis sets was adopted to satisfy the 1 meV/atom convergence criterion.
The cell geometry and crystal basis were fully relaxed using the first-order Methfessel-Paxton smearing method \citep{methfessel1989high} with the width $\sigma$=0.1 eV.
The collinear spin polarization was allowed during the calculations.

One may view a HEA as an undirected graph defined as $G = (V, E)$, where $V$ is the set of nodes representing atoms in three-dimensional coordinate space, and $E$ is the set of edges representing the interaction between any two atoms. Unlike existing literature in graphs, each atom possess the aforementioned attribute \textit{node type} that plays an important role in modeling HEAs; in addition, we adopt \textit{topological structures} to define the closeness of two atoms to account for the periodic pattern in alloys. In particular,
for a HEA supercell consisting of $p$ atoms, we assume that each atom has magnetic interaction with atoms within the closest \textit{topological distance}. For each $i = 1, \ldots, p$, let $(a_{i}, b_{i}, c_{i})$ denote the coordinates of the $i$th atom using its center, and also let $(a_{i}^k, b_{i}^k, c_{i}^k)$ be its coordinates reflected along the $k$th direction, where $k = 1, ..., 6$ indicate up, down, left, right, front and back of the central alloy, respectively. Then the topological distance between atom $i$ and atom $j$, $1\le i,j\le p$ is defined as
\begin{align}
d_{ij}:= {\rm \min} \Big \{& \sqrt {(a_{i}-a_{j})^{2} + (b_{i}-b_{j})^{2} + (c_{i} - c_{j})^{2}},
\\ & \sqrt{(a_{i}-a_{j}^{k})^{2} + (b_{i} -b_{j}^{k})^{2} + (c_{i}-c_{j}^{k})^{2}}: 1 \le k \le 6 \Big \}.
\end{align}

\begin{figure}[!htp]
\centering
\includegraphics [scale = 0.4] {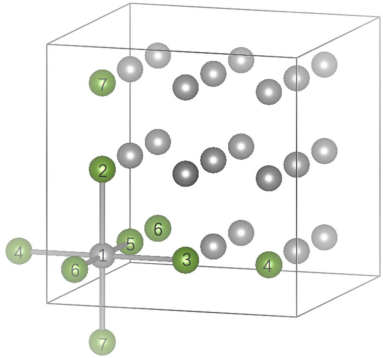}
\caption {A toy example of a HEA supercell.}\label{fig1}
\end{figure}

Figure~\ref{fig1} uses a toy example to demonstrate a supercell and the topological distance. In this example, 27 atoms are evenly distributed in a cube with unit side length, and the radius of each atom is $1/9$. Taking atom 1 (grey) with coordinates $({5}/{6}, {1}/{6}, {1}/{6})$ as the central atom, it has six nearest atoms (green), 2, 3, 4, 5, 6 and 7. Since alloys replicate themselves, atoms 4, 6, 7 appear in the neighboring set of atom 1. The topological distance between atom 1 and any neighboring atom is 1/3. 


\color{black} 

\section{Method} 
\label{sec:method}
We propose a Bayesian conditional autoregressive (CAR) model with graph constraints (CARGO) for magnetic moment interactions in HEA data. In particular, we model the atom interactions using Gaussian Markov random fields through the CAR model, equipped with topological distances and atom types. We formulate a series of constraints that arise in HEA data into graph constraints on the inverse covariance matrix, and propose a constrained Wishart prior in Bayesian CAR. 

We first introduce some notations used throughout this paper. Let $\mathbb{R}^{p}$ and $\mathbb{R}^{p \times q}$ denote the $p$-dimensional Euclidean space and the set of all $p \times q$ real matrices. The set of all natural numbers is denoted by $\mathbb{N}^{p}:=\{1,2,...,p\}$ for $p\in\mathbb{N}$. Denote $\mathbb{S}^{p}$ the space of $p \times p$ symmetric matrices and $\mathbb{S}_{+}^{p}$ ($\mathbb{S}_{++}^{p}$) the set of $p\times p$ symmetric positive semi-definite (definite) matrices. We write $X\succeq 0$ if $X\in\mathbb{S}_{+}^{p}$, and $X \succ 0$ if $X\in\mathbb{S}_{++}^{p}$. For a matrix $X$, $\|X\|_{2}$ denotes the $\ell_{2}$-norm of $X$ which is the largest singular value of $X$, and $\|X\|_{F}$ is the Frobenius norm of $X$ which is equal to $\sqrt{{\rm tr}(XX^{T})}$ with ${\rm tr}(\cdot)$ being the trace of a matrix.
\subsection{Bayesian CAR under graph constraints} \label{sec:model}
Suppose that a HEA has $p$ atoms. Let $x = \{x_{j}\in\mathbb{R}: j = 1, ..., p\} $ be the $p$-variate magnetic moment vector and $s =\{s_j\in\mathbb{N}^{p}: j = 1, ..., p\} $ the corresponding atom type vector. Throughout this article, we focus our analysis on a single sample, enabling the study of transferability across samples, while the developed method easily generalizes to multiple independent samples.

We begin with conditional autoregressive (CAR) models to characterize the conditional dependence structure of atoms, a popular model for spatial data \citep{Wang2013A}. Assume each atom is connected to its nearest $c$ atoms, and the magnetic moment interactions are only determined by the atom types of the two interacting atoms; such constraints stem from the domain knowledge in HEA studies. We assume a conditional Gaussian distribution for the central atom given its neighbors, that is, for any $1 \le i, j\le p$, 
\begin{equation}
p(x_{i} | x_{-i}) = \frac{1}{\sqrt{2 \pi \kappa_{i}}}{\rm exp} \left\{-\frac{1}{2 \kappa_{i}^{2}}(x_{i}-\mu(x_{i}|x_{-i}))^{2}\right\},
\label{898989}
\end{equation}
where
\begin{equation}
\mu(x_{i}|x_{-i}) = \mu_{i} + \sum_{j \in \mathcal{N}_{i}} \beta_{ij}(x_{j}-\mu_{j})
\label{new111}
\end{equation}
is the conditional mean, $\mu = (\mu_{1}, ..., \mu_{p})$ is the marginal mean vector with $\mu_{i} = \mu_{j}$ if $s_{i} = s_{j}$, $\mathcal{N}_{i}$ is the set of neighboring atoms of atom $i$, $\kappa_{i}>0$, and $\beta_{ij}$ represents the spatial coefficient of magnetic moment between atom $i$ and atom $j$ satisfying that
\begin{equation}
\beta_{ij} = \beta_{kl}, \quad \text {if} \ s_{i} = s_{k}\ \text{and} \ s_{j} = s_{l}, \ \text{for\ any}\  1 \le k,l\le p.
\label{2222}
\end{equation}
It is well known that the CAR model leads to the following joint density function \citep{besag1974spatial}
\begin{equation} \label{696969}
p (x|\Sigma^{-1},\mu, s) = \frac{1}{(2\pi)^{p/2} |\Sigma|^{1/2}} {\rm exp} \left\{-\frac{1}{2}(x-\mu)^{T} \Sigma^{-1}(x-\mu)\right\},
\end{equation}
where the $(i,j)$-th entry of $\Sigma^{-1}$ has the explicit form 
\begin{equation}\label{707070}
  \Sigma^{-1}_{ij} =\begin{cases}
            1/\kappa_{i}, & \text{if} \ \ i = j,\\
            -\beta_{ij}/\kappa_{i}, & \text{if} \ \ j \in \mathcal {N}_{i}, \\
               0, & \text{if} \ \ i \notin \mathcal{N}_{j} \text{ and } i \neq j,
            \end{cases}
\end{equation} 
for $i, j = 1, \ldots, p$. The inverse covariance matrix $\Sigma^{-1}$ must satisfy the constraints spelled out in \eqref{2222} and~\eqref{707070} simultaneously. 
The graph constraints can be collectively represented by $\Sigma^{-1}\in N$, where 
\begin{equation}\label{838383}
\begin{aligned}
N = \{X\in & \mathbb{R}^{p\times p}:\; X_{ij} = \Sigma_{ij}^{-1} \text{ satisfies } \eqref{707070}, \\ & \text{ for some } \kappa_{i}>0 \text{ and } \beta_{ij}\in\mathbb{R} \text{ such that } \eqref{2222} \text{ holds for } 1\le i,j\le p.\}.
\end{aligned}
\end{equation}
 In addition to the preceding graph constraints, the inverse covariance matrix $\Sigma^{-1}$ needs to be symmetric and positive definite. The parameter space for $\Sigma^{-1}$ is thus $\widetilde{\Theta}=\mathbb{S}_{++}^{p} \cap N$.
We assume $\mu = 0$ in Model~\eqref{696969} without loss of generality as usually we can center the observations before modeling.  
\subsection{Structured inverse covariance estimation with constrained Wishart prior}
We propose to use a constrained Wishart prior to encode the structures present in the inverse covariance matrix via an indicator function on $\widetilde{\Theta}$. Wishart distributions are widely used as priors for covariance and inverse covariance matrices in the Bayesian literature. We start with a Wishart prior $W(\nu,B)$ on $\Sigma^{-1}$, that is, 
\begin{equation}\label{Wishart_priori}
p(\Sigma^{-1}) = \frac{1}{2^{\nu p/2}}\frac{1}{|B|^{\nu/2}\Gamma_{p}(\nu/2)}|\Sigma^{-1}|^{\frac{\nu-p-1}{2}}{\rm exp}\{-\frac{1}{2}{\rm tr}(B^{-1}\Sigma^{-1})\}, 
\end{equation}
where $B\succeq 0$ is the scale matrix, $\nu$ is the degrees of freedom parameter, and $\Gamma_{p}(\cdot)$ is the multivariate gamma function. Each sample from the Wishart distribution is positive definite with $\nu>p-1$. The unconstrained Wishart distribution has mean $\nu B$, and a larger value of $\nu$ makes the distribution $(\ref{Wishart_priori})$ more widespread while a smaller $\nu$ leads to a more concentrated distribution around $B$. 
Let $\mathbbm{1}_{\Lambda}(\cdot)$ be the indicator function defined on some set $\Lambda\subseteq\mathbbm{R}^{p\times q}$ such that $\mathbbm{1}_{\Lambda}(G) = 1$ if $G \in \Lambda$ and zero otherwise. We then propose to use a constrained prior that respects the graph constraints, that is, $p_{\widetilde{\Theta}}(\Sigma^{-1}) \propto p(\Sigma^{-1})\mathbbm{1}_{\widetilde{\Theta}}(\Sigma^{-1})$.  

Let $S = x^T x$ be the sample covariance matrix; recall that $\mu$ is assumed to be zero. It turns out that the posterior distribution is also a constrained Wishart distribution on $\widetilde{\Theta}$. To see this, we have 
\begin{equation}\label{posterior_distribution}
\begin{aligned}
p_{\widetilde{\Theta}}(\Sigma^{-1}|x,s,\nu)& \propto p(x|\Sigma^{-1},s) \times p_{\widetilde{\Theta}}(\Sigma^{-1}) \\
&\propto |\Sigma^{-1}|^{\frac{1}{2}} {\rm exp} \left\{-\frac{1}{2}{\rm tr}(S\Sigma^{-1})\right\}\cdot|\Sigma^{-1}|^{\frac{\nu-p-1}{2}}{\rm exp}\left\{-\frac{1}{2}{\rm tr}(B^{-1}\Sigma^{-1})\right\} \cdot \mathbbm{1}_{\widetilde{\Theta}}(\Sigma^{-1})\\
& \propto |\Sigma^{-1}|^{\frac{\nu-p}{2}}{\rm exp}\left\{-\frac{1}{2}{\rm tr}((B^{-1}+S)\Sigma^{-1})\right\} \cdot \mathbbm{1}_{\widetilde{\Theta}}(\Sigma^{-1}),
\end{aligned}
\end{equation}
which is the Wishart distribution $W(\nu + 1,(B^{-1}+S)^{-1})$ constrained on $\widetilde{\Theta}$.

We are interested in deriving the maximum a posterior (MAP) estimate $\Sigma^{-1}_{\rm MAP}$, which amounts to the following constrained minimization problem
\begin{equation}\label{Problem_original}
\Sigma^{-1}_{\rm MAP} = \underset{X\in \widetilde{\Theta}}{\arg \min}\{ -{\rm (\nu-p)log}({\rm det}\,X)+{\rm tr}((B^{-1}+S)X)\} =:\underset{X\in \widetilde{\Theta}}{\arg \min} \; \{F(X)\}.
\end{equation}
Note that the function $F(X)$ depending on the log-determinant function is strongly convex.
Let ${\rm lev}(F,\alpha):= \{X\in\mathbb{S}_{++}^{p}:F(X)\le\alpha\}$ be the level set of $F(X)$. The following lemma illustrates the level boundedness of the objective function $F(X)$.
\begin{lemma}\label{level_boundedness_FX}
For $F$ defined in (\ref{Problem_original}) and any $\alpha > {\rm inf}\,F$, there exist $0 < C_{1}\le C_{2}$, such that ${\rm lev}(F,\alpha)\subseteq \{X\in\mathbb{S}_{++}^{p}: C_{1} I \preceq\|X\|_{2}\preceq C_{2} I\}$.
\end{lemma}
\begin{proof}
For any $X\in {\rm lev}\,(F,\alpha)$, since ${\rm log}({\rm det}\,X)\le p\,{\rm log}\|X\|_{2}$, there holds
\begin{align}
\alpha + (\nu-p)p{\rm log}\|X\|_{2} &\ge \alpha + (\nu-p){\rm log}({\rm det}\,X) \\
&\ge {\rm tr}((B^{-1} + S)X)\\
&\ge \|(B^{-1}+S)X\|_{2} \\
&\ge \sigma_{\min}(B^{-1}+S)\|X\|_{2}, 
\end{align} 
where $\sigma_{\min}(\cdot)$ denotes the smallest eigenvalue of a matrix. Let $\sigma_{m}:=\sigma_{\min}(B^{-1}+S)$, which is positive as $B$ is positive definite and $S$ is semi-positive definite. Define $g(x):=\alpha + p(\nu-p)\,{\rm log}x-\sigma_{m}x$ for $x > 0$. By direct calculation of its first and second derivatives, the function $g(x)$ is strictly concave with its global maximum attained at $x = {p(\nu-p)}/{\sigma_{m}}$. Noting that $g(x)\rightarrow -\infty$ if $x\rightarrow 0+$ and $x\rightarrow +\infty$, there exist $0< 
c\le C_{2}$ such that ${\rm lev}(g,0) = \{x:c\le x\le C_{2}\}$. This yields the upper bound, that is, ${\rm lev}(F,\alpha)\subseteq \{X\in\mathbb{S}_{++}^{p}: X\preceq C_{2} I\}$.

For the lower bound, since ${\rm tr}((B^{-1} + S)X)\ge 0$, there holds
$$
\alpha > -(\nu-p){\rm log\,det}\,(X) > -(\nu-p)\{\log\sigma_{\min}(X) - (p-1)\log\sigma_{\max}(X)\},
$$
where $\sigma_{\max}(\cdot)$ is the largest eigenvalue of a matrix. Hence, a lower bound for $\sigma_{\min}(X)$ is $C_{1} = e^{-\alpha/(\nu-p)}C_{2}^{-(p-1)}$.
This completes the proof. 
\end{proof}

According to Lemma \ref{level_boundedness_FX}, if we choose a small number $\epsilon\le C_{1}$ such that the global minimizer $X^{\star}$ of Problem (\ref{Problem_original}) satisfies $X^{\star}\succeq \epsilon I$, then Problem (\ref{Problem_original}) can be recast into its equivalent counterpart as follows
\begin{equation}\label{new222}
{\rm \min}\{F(X) :X-\epsilon I \in\mathbb{S}_{+}^{p}, \;X \in N \}.
\end{equation}
The constraint set $\{X:X-\epsilon I \in\mathbb{S}_{+}^{p}, \;X \in N \}$ is closed, which along with the strong convexity and continuity of $F$ ensures the existence of the optimal solution of Problem (\ref{new222}). This is presented in the following result (proof omitted) in view of the Weierstrass theorem \citep{Apostol_mathematics}. 
\begin{theorem}\label{existence_of_minimizer}
There exists a unique minimizer for Problem (\ref{new222}). 
\end{theorem}
\section{Modified Gauss-Seidel algorithm and convergence analysis} 
\label{sec:algorithm.convergence}

\newcommand{\MAP}{\mathrm{MAP}}

In this section, we propose a penalty decomposition method via a modified Gauss-Seidel scheme for Problem~\eqref{new222} to estimate $\Sigma_{\MAP}^{-1}$. We derive closed-form expressions using proximity operators and establish algorithmic guarantees, enabling efficient posterior calculation. 

The parameter space entails both graph constraints and positive definiteness for $\Sigma^{-1}$. We disentangle these two constraints using penalty decomposition. Letting
$
M = \{X\in\mathbb{S}^{p}:X-\epsilon I\succeq 0\},
$
we rewrite Problem (\ref{new222}) as
\begin{equation}\label{767676}
\mathop{\rm \,\min} \,\left \{ \, F(X):\,X=Y,\, X\in M, \, Y\in N\right \},
\end{equation}
which is further turned to the following penalty decomposition problem
\begin{equation}\label{777777}
\mathop{\rm \,\min} \,\left \{ \, F(X)+\frac{\gamma}{2}\|X-Y\|_{F}^{2}:\,(X,Y)\in M\times N\right \} =: \mathop{\rm \,\min} \,\left \{ \, P_{\gamma}(X,Y):\,(X,Y)\in M\times N\right \},
\end{equation}
where $\gamma>0$ is a penalty parameter, and $P_{\gamma}$ maps $(X,Y)\in M\times N$ to 
\begin{equation}\label{new666}
P_{\gamma}(X,Y):= -(\nu-p){\rm log}({\rm det}\,X)+{\rm tr}((B^{-1}+S)X)+\frac{\gamma}{2}\|X-Y\|_{F}^{2}.
\end{equation}

The proposed optimization algorithm is outlined by a two-loop iterative scheme, detailed in the next two sections. The inner loop executes a proximal modification of the Gauss-Seidel method for solving the penalty decomposition problem in (\ref{777777}) for given $\gamma$, while the outer loop increases the penalty parameter $\gamma$ to push the output of inner loop to the desired solution of Problem (\ref{new222}). 

\subsection{Proximal modification of the Gauss-Seidel method for Problem \eqref{777777}}
The following theorem guarantees the inner loop Problem~\eqref{777777} has a unique solution for any given $\gamma$. 
\begin{theorem}\label{level_boundedness_penalty_problem}
For any $\gamma>0$, there exists a unique minimizer for the problem defined in (\ref{777777}).
\end{theorem}
\begin{proof}
Since $P_{\gamma}$ is a strongly convex function from $\mathbb{S}_{++}^{p}$ to $\mathbb{R}$, it is sufficient to show its level boundedness. For any $\alpha > {\rm inf}\,P_{\gamma}$ and  $(X,Y)\in {\rm lev}(P_{\gamma},\alpha)$, there holds \begin{align}
    \alpha & \ge P_{\gamma}(X,Y) = -(\nu-p){\rm log}({\rm det}\,X)+{\rm tr}((B^{-1}+S)X)+\frac{\gamma}{2}\|X-Y\|_{F}^{2} \\
    & \ge-(\nu-p){\rm log}({\rm det}\,X)+{\rm tr}((B^{-1}+S)X).
\end{align} According to lemma \ref{level_boundedness_FX}, $\|X\|_{2}$ can be bounded by some constant $C_{2}$. For $Y$, noting ${\rm tr}(B^{-1}+S)>0$, we have $\alpha\ge P_{\gamma}(X,Y)\ge -(\nu-p){\rm log}({\rm det}\, X)+\frac{\gamma}{2}\|X-Y\|_{2}^{2}$. Using the triangle inequality under the $\ell_{2}$-norm, we arrive at $\sqrt{\frac{\gamma}{2}}\|Y\|_{2}\le \sqrt{\frac{\gamma}{2}}\|X\|_{2} + \sqrt{\frac{\gamma}{2}}\|Y-X\|_{2}\le \sqrt{\frac{\gamma}{2}}\|X\|_{2} + \sqrt{p(\nu-p){\rm log}(\|X\|_{2})+\alpha}$, which yields $\|Y\|_{2}\le C_{2} + \sqrt{\frac{\gamma}{2}}\sqrt{p(\nu-p){\rm log}\,C_{2}+\alpha}.$ This shows the level boundedness of ${\rm lev}(P_{\gamma},\alpha)$ and completes the proof.
\end{proof}

Let $\phi_{\Lambda}$ be a function on a set $\Lambda\subseteq\mathbb{R}^{p\times q}$, defined at $\Omega \in \Lambda$ as
\begin{equation}\label{0_infty_function}
\phi_{\Lambda}(\Omega) := \left\{
\begin{aligned}
&0, \quad \quad \Omega\in\Lambda
\\
&+\infty, \ \Omega\notin\Lambda.
\end{aligned}
\right.
\end{equation}
Define $G_{\gamma}(X,Y):= P_{\gamma}(X,Y) + \phi_{M}(X) +\phi_{N}(Y)$. The optimization problem in (\ref{777777}) is turned to
\begin{equation}\label{new555}
{\rm \min}\{G_{\gamma}(X,Y):(X,Y)\in\mathbb{R}^{p\times p}\times \mathbb{R}^{p\times p}\},
\end{equation} 
which is a convex optimization problem; this is because the constraint sets $M$ and $N$ are convex by simple matrix manipulations, and $P_{\gamma}(X,Y)$ in (\ref{new666}) is also convex on $\{(X,Y): (X,Y)\in M\times N\}$. The function $P_{\gamma}(X,Y)$ satisfies the Kurdyka-Lojasiewicz property since it has a log-exp structure which belongs to tame functions \citep{H2010Proximal}. Tame functions \citep{Bolte2009Tame} encompass a large class of functions, including piecewise polynomial, analytic functions, and log-exp structure, and satisfy the Kurdyka-Lojasiewicz inequality that is critical in analyzing convex/nonconvex optimization problems.

For any $l\in\mathbb{N}$ with initial point $(X_{0},Y_{0})\in M\times N$, the characterizations above yield a proximal modification of the Gauss-Seidel method for minimizing $G_{\gamma}(\cdot, \cdot)$ by applying the following alternating minimization procedure
\begin{equation}\label{new444}
\left\{
\begin{aligned}
X_{l+1}&= {\rm arg}\mathop{\rm \,\min} \left \{ \, G_{\gamma}(X,Y_{l}) +\frac{1}{2}\|X-X_{l}\|_{F}^{2}:X\in \mathbb{R}^{p\times p} \right \},
\\
Y_{l+1}&= {\rm arg}\mathop{\rm \,\min} \left \{ \, G_{\gamma}(X_{l+1},Y)+\frac{1}{2}\|Y-Y_{l}\|_{F}^{2}:Y\in \mathbb{R}^{p\times p} \right \}.
\end{aligned}
\right.
\end{equation}

This iterative scheme is a class of alternating minimization algorithms with ``costs to move" terms $\|X-X_{l}\|_{F}^{2}$ and $\|Y-Y_{l}\|_{F}^{2}$. \citep{2008Alternating} illustrates that these terms play a crucial role in the convergence of the algorithm, especially for preventing divergent sequences in the weak coupling case. 
We obtain the following convergence results for  $\{(X_{l},Y_{l}):l\in\mathbb{N}\}$.
\begin{theorem}\label{Theorem3}
The sequence $\{(X_{l},Y_{l}):l\in\mathbb{N}\}$ generated by (\ref{new444}) with some initial point $(X_{0},Y_{0})\in M\times N$ is bounded and converges to the global minimizer $(\overline{X},\overline{Y})$ of Problem $(\ref{new555})$.
Moreover, the sequence $\{(X_{l},Y_{l}):l\in\mathbb{N}\}$ has a finite length, i.e., $\|(X_{l+1},Y_{l+1})-(X_{l},Y_{l})\|_{F}^{2}<+\infty$.
\end{theorem}
\begin{proof} 
For any $\gamma>0$, we first show that the sequence $\{(X_{l},Y_{l}):l\in\mathbb{N}\}$ generated by (\ref{new444}) with initial point $(X_{0},Y_{0}) \in M\times N$ is bounded. For any $l\in\mathbb{N}$, there holds
\begin{equation}\label{supp_111}
G_{\gamma}(X_{l+1},Y_{l}) + \frac{1}{2}\|X_{l+1}-X_{l}\|_{F}^{2}\le G_{\gamma}(X_{l},Y_{l}),
\end{equation} 
\begin{equation}\label{supp_222}
G_{\gamma}(X_{l+1},Y_{l+1}) + \frac{1}{2}\|Y_{l+1}-Y_{l}\|_{F}^{2}\le G_{\gamma}(X_{l+1},Y_{l}).
\end{equation}
Combining $(\ref{supp_111})$ and $(\ref{supp_222})$ yields
\begin{equation}\label{supp_333}
G_{\gamma}(X_{l+1},Y_{l+1})+\frac{1}{2}\|X_{l+1}-X_{l}\|_{F}^{2}+\frac{1}{2}\|Y_{l+1}-Y_{l}\|_{F}^{2} \le G_{\gamma}(X_{l},Y_{l}). 
\end{equation}
Then we have $G_{\gamma}(X_{l+1},Y_{l+1}) \le G_{\gamma}(X_{l},Y_{l}) \le G_{\gamma}(X_{0},Y_{0})$, where $G_{\gamma}(X_{0},Y_{0})$ is some constant. This combined with the level boundedness of $G_{\gamma}$ by the proof of Theorem \ref{level_boundedness_penalty_problem} leads to the boundedness of the sequence $\{(X_{l},Y_{l}):l\in\mathbb{N}\}$.

Note that $G_{\gamma}(\cdot, \cdot)$ for $\gamma>0$ is a lower semicontinuous function and satisfies the Kurdyka-Lojasiewicz property. One can also check that the Lipschitz constant of the derivative of $\frac{\gamma}{2}\|X-Y\|_{2}$ is $2\gamma$. According to Theorem 6.2 of \citep{attouch2013convergence}, it follows that $\{(X_{l},Y_{l}):l\in\mathbb{N}\}$ converges to some critical point $(\overline{X},\overline{Y})$ of $G_{\gamma}(X,Y)$ and has a finite length, i.e., $\sum_{k}\|(X^{k+1},Y^{k+1})-(X^{k},Y^{k})\|_{F}^{2}<+\infty$. Since $G_{\gamma}(X,Y)$ is convex, it has at most one critical point. Hence, $\{(X_{l},Y_{l}):l\in\mathbb{N}\}$ converges to the global minimizer of Problem (\ref{new555}).
\end{proof}

We next turn to deriving closed-form expressions for the iterative scheme in (\ref{new444}) by utilizing proximity operators \citep{Bauschke2011Convex}, enabling fast computation. For a proper, lower semi-continuous function $f:\mathbb{R}^{p\times p}\rightarrow \mathbb{R}$, the proximity operator of $f$ at $Z\in \mathbb{R}^{p\times p}$ is defined by 
\begin{equation} \label{eq:prox.def}
{\rm prox}_{f}(Z):={\rm arg\,\min}\{f(U)+\frac{1}{2}\|U-Z\|_{F}^{2}:U\in\mathbb{R}^{p\times p}\}.
\end{equation}
Denote
\begin{equation}\label{909090}
L(X) = -(\nu-p){\rm log}({\rm det}X) + {\rm tr}((B^{-1}+S)X)+\phi_{M}(X),
\end{equation}
and rewrite (\ref{new444}) as
\begin{equation}\label{new888}
\left\{
\begin{aligned}
X_{l+1}&={\rm prox}_{\frac{1}{\gamma+1}L}\left(\frac{1}{\gamma+1}X_{l}+\frac{\gamma}{\gamma+1} Y_{l}\right),
\\
Y_{l+1}&={\rm prox}_{\phi_{N}}\left(\frac{\gamma}{\gamma+1}X_{l+1}+\frac{1}{\gamma+1} Y_{l}\right).
\end{aligned}
\right.
\end{equation} 
Define ${\rm diag}(z)$ for $z\in \mathbb{R}^{p}$ as the $p\times p$ diagonal matrix whose diagonal entries are $z = (z_{1},...,z_{p})$. We then obtain closed-form expressions for the the two proximity operators in (\ref{new888}) in the following proposition. 
\begin{proposition}\label{Proposition1}
Suppose $\alpha>0$. Let $L$ and $N$ be defined by $(\ref{909090})$ and $(\ref{838383})$, respectively. Then the following results hold. \\
(i)\;For any $Z\in \mathbb{S}^{p}$, let $V{\rm diag}(\sigma) V^{T}$ be the eigenvalue decomposition of $\frac{1}{\alpha}Z-(B^{-1}+S)$, where $\sigma = (\sigma_{1},...,\sigma_{p})$ collects all the $p$ eigenvalues of $\frac{1}{\alpha} Z - (B^{-1}+S)$, and $V^{T}V = I$. Then, we have
\begin{equation}
{\rm prox}_{\alpha L}(Z)= V{\rm diag}(z)V^{T},
\end{equation}
where $z = (z_1, \ldots, z_p)$ with $$z_{i}:= \max \left\{\frac{\alpha}{2}\left(\sigma_{i}+\sqrt{\sigma_{i}^{2}+\frac{4(\nu-p)}{\alpha}}\right),\epsilon\right\}.$$\\
(ii)\;For any $Z\in \mathbb{S}^{p}$, let $Z_{ij}$ be the (i,j)-th entry of $Z$ for any $i,j = 1,...,p$, then
\begin{equation}
\begin{aligned}
{\rm prox}_{\phi_{N}}(Z) = \Bigg\{ U \in\mathbb{R}^{p\times p}:U_{ii} &= \frac{\sum_{j = 1}^{p}Z_{jj}\cdot\mathbbm{1}(s_{j} = s_{i})}{\sum_{j=1}^{p}\mathbbm{1}(s_{j} = s_{i})}, \\
U_{ij} &= \frac{\sum_{k,l = 1}^{p}Z_{kl}\cdot\mathbbm{1}(s_{k} = s_{i}, s_{l} = s_{j})}{\sum_{k,l=1}^{p}\mathbbm{1}(s_{k} = s_{i}, s_{l} = s_{j})}, \ \text{if}\ j\in\mathcal{N}_{i},\\
U_{ij}& = 0,  \ \text{if}\ j\notin\mathcal{N}_{i} \text{ and } i \neq j.\Bigg\}, 
\end{aligned}
\end{equation}
where $\mathbbm{1}(\omega)$ is the indicator function of $\omega$ on $\mathbb{R}$.
\end{proposition}
\begin{proof}
Substituting the function $L$ defined in $(\ref{909090})$ into~\eqref{eq:prox.def} yields
\begin{equation}\label{proximity_operator_L}
\begin{aligned}
& \;{\rm prox}_{\alpha L}(Z) = {\rm arg\,\min}\left\{\alpha L(X) + \frac{1}{2}\|X-F\|_{F}^{2}:X\in M\right\}\\
= &\; {\rm arg\,\min}\left\{-\alpha(\nu-p){\rm log}({\rm det}X)+\alpha{\rm tr}((B^{-1}+S)X) + \frac{1}{2}\|X-Z\|_{F}^{2}:X\in M\right\}.
\end{aligned}
\end{equation}
The first-order condition of $(\ref{proximity_operator_L})$ satisfies
\begin{equation}\label{first-order}
-(\nu-p)X^{-1} + \frac{1}{\alpha} X = \frac{1}{\alpha} Z - (B^{-1}+S).
\end{equation}
Let $V{\rm diag}(\sigma) V^{T}$ be the eigenvalue decomposition of $\frac{1}{\alpha}Z-(B^{-1}+S)$, where $\sigma = (\sigma_{1},...,\sigma_{p})$ collects all the $p$ eigenvalues of $\frac{1}{\alpha} Z - (B^{-1}+S)$, and $V^{T}V = I$. Multiplying (\ref{first-order}) by $V^{T}$ on the left and by $V$ on the right, we obtain
\begin{equation}\label{equation_diagonal}
-(\nu-p)\hat{X}^{-1} + \frac{1}{\alpha} \hat{X} = {\rm diag}(\sigma),
\end{equation}
where $\hat{X} = V^{T}XV$ is a diagonal matrix since the left-hand side of (\ref{equation_diagonal}) is diagonal. Hence, $\hat{X}_{ii}$ is a positive root of the quadratic function $\frac{1}{\alpha}\hat{X}_{ii}^{2}-\sigma_{i}\hat{X}_{ii}-(\nu-p)$, which, combined with $X-\epsilon I \succeq 0$ as $X\in M$, yields that $$\hat{X}_{ii} = \max \left\{\frac{\alpha}{2}\left(\sigma_{i}+\sqrt{\sigma_{i}^{2}+\frac{4(\nu-p)}{\alpha}}\right),\epsilon\right\}.
$$
This completes the proof of (i). 

We next prove (ii). 
We can see that ${\rm prox}_{\phi_{N}}(Z)$ is the projection of $Z$ on the constraint set $N$. First notice that for any $i,j = 1,...,p$, there holds $U_{ij} = 0$ if $j\notin\mathcal{N}_{i}$ for $ i\neq j$. Denote the index set of non-zero elements of $U\in N$ by $\Lambda:=\{(i,j):U\in N,\ U_{ij}\neq 0\} \subset \{(i, j): i = j \text{ or } j \in \mathcal{N}_i, \; i, j = 1, \ldots, p\}$. We next calculate the value of $U$ on its non-zero position set $\Lambda$. The Frobenius norm of $U-Z$ can be written as
$
\|U-Z\|_{F}^{2} = \sum_{(i,j)\in\Lambda}(U_{ij}-Z_{ij})^{2}+\sum_{(i,j)\in\Lambda^{c}}(Z_{ij})^{2}.
$
Then, to minimize $\|U-Z\|_{F}^{2}$ is equivalent to minimize
\begin{equation}\label{939393}
\sum_{(i,j)\in\Lambda}(U_{ij}-Z_{ij})^{2}.
\end{equation}
According to the definition of $N$, $\text{for any} \  i,j = 1,...,p$ such that $i\neq j$, we have 
$$
U_{ii} = U_{jj},\  \text{if}\ s_{i} = s_{j}  
$$
and
$$
U_{ij} = U_{ji} = U_{kl} = U_{lk},\  \text{if}\ j\in \mathcal{N}_{i}, l\in \mathcal{N}_{k}, \ s_{i} = s_{k}, s_{j} = s_{l}.
$$
By computing the stationary points of the quadratic function~\eqref{939393} using the first-order conditions for optimality, we arrive at
$$
U_{ii} = \frac{\sum_{j = 1}^{p}Z_{jj}\cdot\mathbbm{1}(s_{j} = s_{i})}{\sum_{j=1}^{p}\mathbbm{1}(s_{j} = s_{i})},
$$
$$
U_{ij} = \frac{\sum_{k,l = 1}^{p}Z_{kl}\cdot\mathbbm{1}(s_{k} = s_{i}, s_{l} = s_{j})}{\sum_{k,l=1}^{p}\mathbbm{1}(s_{k} = s_{i}, s_{l} = s_{j})},
$$
for any $i,j = 1,...,p$. This completes the proof. 
\end{proof}

\subsection{Penalty decomposition proximal modification of the Gauss-Seidel method} \label{sec:algorithm} 
We are now in a position to present the penalty decomposition proximal modification of the Gauss-Seidel method to obtain $\Sigma_{\MAP}^{-1}$ by solving the optimization problem in (\ref{new222}).
The pseudo-code is provided in Algorithm~\ref{alg:Framwork}. 
\begin{algorithm}[htb] 
\caption{Penalty decomposition proximal modification of the Gauss-Seidel algorithm for solving Sparse inverse covariance estimation with graph constraints.} 
\label{alg:Framwork} 
\begin{algorithmic}[1] 
\Require 
Graph constraints $N$, a positive decreasing sequence $\{\epsilon^{k}:k\in \mathbb{N}\}$ with $\epsilon^{k}\to 0_{+}$, initial penalty parameter $\gamma^{0}>0$, $k_{\max}>0$, $\epsilon>0$, $\nu>p-1$, $B\succeq 0$, $\eta>1$.\\ 
\textbf{Set}: \quad \quad  $k=0,\ \ X^{0} = I$, $Y^{0}\in\mathbb{S}_{++}^{p}$.\\
\textbf{While} \quad $k<k_{\max}$\\
\qquad \qquad \textbf{Set:}  \quad \  \ $l = 0$, $X_{0}^{k} = X_{-1}^{k}=X^{k}$, $Y_{0}^{k} = Y_{-1}^{k}=Y^{k}$\\
\qquad \qquad \textbf{While} \quad$l = = 0$ or ${\rm max}\{\| (X_{l}^{k}-X_{l-1}^{k})+\gamma^{k}(Y_{l}^{k}-Y_{l-1}^{k})\|_{F},\|Y_{l}^{k}-Y_{l-1}^{k}\|_{F}\}>\epsilon^{k}$\\
$$
\left\{
\begin{aligned}
X_{l+1}^{k}&= {\rm arg}\mathop{\rm \,\min} \left \{ \, G_{\gamma^{k}}(X,Y_{l}^{k}) +\frac{1}{2}\|X-X_{l}^{k}\|_{F}^{2}:X\in \mathbb{R}^{p\times p} \right \},
\\
Y_{l+1}^{k}&= {\rm arg}\mathop{\rm \,\min} \left \{ \, G_{\gamma^{k}}(X_{l+1}^{k},Y)+\frac{1}{2}\|Y-Y_{l}^{k}\|_{F}^{2}:Y\in \mathbb{R}^{p\times p} \right \},
\end{aligned}
\right.
$$
\qquad \qquad \qquad \qquad $l:=l +1$,\\
\qquad \qquad \textbf{end}\\
\qquad \qquad$(X^{k+1},Y^{k+1}):=(X_{l}^{k},Y_{l}^{k})$,\\
\qquad \qquad$\gamma^{k+1}:=\eta\gamma^{k}$, \\
\qquad \qquad$k:=k+1$.\\
\textbf{end}
\end{algorithmic} 
\end{algorithm}
Note that Algorithm~\ref{alg:Framwork} makes use of a verifiable stopping rule for the inner loop, i.e., ${\rm max}\{\|(X_{l}^{k}-X_{l-1}^{k})+\gamma^{k}(Y_{l}^{k}-Y_{l-1}^{k})\|_{F},\|Y_{l}^{k}-Y_{l-1}^{k}\|_{F}\}\le\epsilon^{k}$. The following theorem analyzes convergence properties of Algorithm~\ref{alg:Framwork}, which converges very fast confirmed by both simulations and applications to real HEA data.
\begin{theorem}\label{Theorem5}
Let $\{(X^{k},Y^{k}):k\in\mathbb{N}\}$ be generated by Algorithm 1. Then the following statements hold:\\
(i) The sequence $\{(X^{k},Y^{k}):k\in\mathbb{N}\}$ is bounded.\\
(ii) $\lim_{k \rightarrow \infty} \|X^{k}-Y^{k}\|_{F} = 0.$\\
(iii) Let $(X^{\star},Y^{\star})$ be any accumulation point of $\{(X^{k},Y^{k}):k\in\mathbb{N}\}$. Then, $X^{\star} = Y^{\star}$ and $X^{\star}$ is the global minimizer of problem $(\ref{new222})$.
\end{theorem}
\begin{proof}
We first prove (i). Theorem \ref{Theorem3} indicates that $(X^{k},Y^{k})$ is a global minimizer of Problem (\ref{new555}) with penalty parameter $\gamma^{k}$ for any $k\in\mathbb{N}$. For $(\epsilon I, \epsilon I)\in M\times N$, there holds 
\begin{equation}\label{supp_444}
F(X^{k}) + \frac{\gamma^{k-1}}{2}\|X^{k}-Y^{k}\|_{F}^{2}\le F(\epsilon I).
\end{equation}
Then we have $F(X^{k})\le F(\epsilon I)$. In view of the level boundedness of the function $F$ by Lemma \ref{level_boundedness_FX}, it follows that $\{X^{k},k\in\mathbb{N}\}$ is bounded. From the iterative scheme of the inner loop in Algorithm~\ref{alg:Framwork}, we have 
\begin{equation}\label{supp_666}
G_{\gamma^{k-1}}(X^{k},Y^{k})\le G_{\gamma^{k-1}}(X^{k+1},Y^{k+1}) \le
G_{\gamma^{k}}(X^{k+1},Y^{k+1})\le
G_{\gamma^{k}}(X^{k},Y^{k}),
\end{equation}
which gives $G_{\gamma^{k}}(X^{k},Y^{k}) - G_{\gamma^{k-1}}(X^{k},Y^{k}) \geq G_{\gamma^{k}}(X^{k+1},Y^{k+1}) - G_{\gamma^{k-1}}(X^{k+1},Y^{k+1})$. Hence, 
we obtain
$$
(\gamma^{k}-\gamma^{k-1})\|X^{k+1}-Y^{k+1}\|_{F}^{2}\le (\gamma^{k}-\gamma^{k-1})\|X^{k}-Y^{k}\|_{F}^{2}.
$$
Since $\gamma^{k}\ge \gamma^{k-1}$, it holds that
$$
0\le \|X^{k+1}-Y^{k+1}\|_{F}^{2}\le \|X^{k}-Y^{k}\|_{F}^{2},
$$
which means the sequence $\{\|X^{k}-Y^{k}\|_{F}^{2}:k\in\mathbb{N}\}$ is convergent. This shows that $\{(X^{k},Y^{k}),k\in\mathbb{N}\}$ is bounded.  

For (ii), first note that there holds $$\frac{\gamma^{k-1}}{2}\|X^{k}-Y^{k}\|_{F}^{2}\le F(\epsilon I) - F(X^{k})\le F(\epsilon I)- {\rm inf}\,F,$$
according to $(\ref{supp_444})$. Since $\gamma^{k}\rightarrow +\infty$ as $k\rightarrow +\infty$, we have $\mathop{\rm lim}\limits_{k\rightarrow \infty}\|X^{k}-Y^{k}\|_{F} = 0$.

We finally prove (iii). Let $\gamma>\gamma^{0}>0$. By (\ref{new888}), we have $X_{l} + Y_{l} - (\gamma + 1)X_{l}\in \partial L(X_{l})$, which induces that $(X_{l}-X_{l+1})+\gamma(Y_{l}-Y_{l+1})\in\partial F(X_{l}) +\gamma(X_{l+1}-Y_{l+1}) + \ell_{M}(X_{l+1})$. Together with $Y_{l+1} = {\rm prox}_{\ell_{N}}\left(\frac{\gamma}{\gamma + 1}X_{l+1}+\frac{1}{\gamma + 1}Y_{l}\right)$, it follows that $(Y_{l}-Y_{l+1})\in\gamma(Y_{l+1}-X_{l+1}) + \partial_{\ell_{N}}(Y_{l+1})$. Since the objective function in (\ref{777777}) is strong convex with constant $\gamma^{0}$, there holds
\begin{equation}\label{supp_555}
\left\|
\left(
  \begin{array}{cc}
   X^{k+1}-\hat{X}^{k} \\
   Y^{k+1}-\hat{Y}^{k}\\
  \end{array}
\right)
\right\|_{F}\le \frac{\sqrt{2}\epsilon^{k}}{\gamma^{0}},
\end{equation}
where  
\begin{equation}
(\hat{X}^{k},\hat{Y}^{k}) = \mathop{\rm arg\,\min} \,\left \{ \, F(X)+\frac{\gamma^{k}}{2}\|X-Y\|_{F}^{2}:\,(X,Y)\in M\times N\right \}.
\end{equation} 
By passing to a subsequence if necessary, we assume $X^{k}\rightarrow X^{\star}$ and $Y^{k}\rightarrow Y^{\star}$. Due to $\epsilon^{k}\rightarrow 0$, we also have $\hat{X}^{k}\rightarrow X^{\star}$ and $\hat{Y}^{k}\rightarrow Y^{\star}$. It follows from the classical results of penalty methods \citep{JorgeNocedal2006Numerical} that as $\{\gamma^{k}>0:k\in\mathbb{N}\}$ goes to infinity, the accumulation point $({X}^{\star},{Y}^{\star})$ of $\{(\hat{X}^{k},\hat{Y}^{k}):k\in\mathbb{N}\}$ is the global minimizer of Problem (\ref{767676}), that is, $X^{\star}$ is the global minimizer of Problem (\ref{new222}). 
\end{proof} 

In Algorithm~\ref{alg:Framwork}, the diagonal elements of $Y^{0}$ are 1, and the pattern of non-zero elements at non-diagonal positions of $Y^{0}$ is same as $N$ with a relatively small number guaranteed the positive definiteness of $Y_{0}$. We use $\nu = p + 1$ and set $k_{\max} = 35$, $\eta = 2$, $\epsilon = e^{-4}$, $\epsilon_{k} = 2^{-k}$ in Algorithm~\ref{alg:Framwork} when implementing the proposed method, unless stated otherwise. 

\section {Simulation}\label{sec:simulation} 
In this section, we carry out simulations to evaluate CARGO relative to competing methods in terms of parameter estimation, and assess convergence of the proposed optimization algorithm. 


For each $p$, the generated data consist of a $p$-dimension atom type vector $s$ and observations $x$ from $\mathcal{N}(0,\Sigma^{-1})$. We consider three node types labeled as $\{1, 2, 3\}$. Each entry $s_i$ of $s$ is randomly drawn from a uniform distribution on $\{1,2,3\}$. The inverse covariance matrix $\Sigma^{-1}$ relies on $\beta_{ij}$ and $\kappa_i$ as in (\ref{707070}) for $i,j = 1,...,p$. The coefficients $\beta_{ij}$ are specified by $s$ according to the values in Table \ref{table22}.
\begin{table}[htbp]
         \centering
         \caption{Specification of model parameters $\beta_{ij}$ with three node types.}
        \begin{tabular}{llll}
\toprule
\diagbox{$s_{j}$}{$\beta_{ij}$}{$s_{i}$} & $1$ & $2$ & $3$ \\ 
\cmidrule{1-1} \cmidrule{2-4}
$1$&$\beta_{1} = 0.0500 $&$\beta_{4}$&$\beta_{5}$ \\
$2$&
$\beta_{4} = 0.1050$&$\beta_{2} = 0.1000$&$\beta_{6}$ \\
$3$
&$\beta_{5} = -0.0625$&$\beta_{6} = -0.1250$&$\beta_{3} = -0.1000$ \\
           \bottomrule
           \end{tabular}
           \vspace{0.05in}
        \label{table22}
    \end{table}
We use homogeneous $\kappa_{i} = \kappa >0$ for $i = 1,2, ..., p$. We focus on sparse $\Sigma^{-1}$ without encoding topological distances nor the topological neighboring sets in the data generating scheme to give competing methods an advantage---unlike CARGO they do not incorporate such structures by design. As such, we randomly choose about 10$\%$ of total elements of $\Sigma^{-1}$ as non-zero, and assign atom type-dependent $\beta_{ij}$ to non-zero positions of $\Sigma^{-1}$. We further enlarge the diagonal entries until the matrix becomes positive definite. 

We vary the dimension by considering $p = \{20, 50, 100\}$. Each simulation is replicated $K = 50$ times. For each method, we calculate the mean squared error (MSE) to evaluate the accuracy: 
$$
{\rm MSE}_{l} = \frac{\sum_{k = 1}^{K}(\hat{\beta}_{l}^{(k)}-\beta_{l})^{2}}{K},
$$
where $\hat{\beta}_{l}^{(k)}$ is the estimate of $\beta_{l}$ in the $k$th replication, for $l = 1, \ldots, 6$.



For the proposed method CARGO, we implement Algorithm~\ref{alg:Framwork}, following the specification of input parameters in Section~\ref{sec:algorithm}. We compare CARGO to another two methods: graphical Lasso \citep{Friedman_2007}, or gLasso, and a recent Penalized Likelihood method with Tikhonov regularization under an $\ell_{0}$ constraint \citep{Liu_2019}, or PLT$_0$. 
Graphical lasso estimates $\Sigma^{-1}$ by
\begin{equation}\label{Graphical_lasso_model}
\min\left\{ -\log(\,\det\,X)+{\rm tr}(SX)+\lambda \|X\|_{1}\right \}
\end{equation} 
over non-negative definite matrices $X$, where $\lambda>0$ is the regularization parameter. 
The implementation of gLasso is fast based on iteratively solving $p$ Lasso problems. PLT$_0$ estimates sparse inverse covariance matrices by solving the following $\ell_{0}$-norm constrained Tikhonov regularized negative log-likelihood minimization problem
\begin{equation}\label{PDFPPA_model}
\min\left \{ -\log(\,\det\,X)+{\rm tr}(SX)+\frac{\lambda}{2}\|X\|_{F}^{2} : \, X\in\mathbb{S}_{++}^{p}, \, \|X\|_{0}\le s^*\right \},
\end{equation}
where $\lambda>0$ is the regularization parameter and $s^*$ is a given even number to specify the sparsity level. 
Tuning of the regularization parameter $\lambda$ in PLT$_0$ follows the recommendations by the authors. We give PLT$_0$ an advantage by setting the sparsity parameter $s^*$ in
(\ref{PDFPPA_model}) to the truth; likewise, we tune the regularization parameter $\lambda$ in (\ref{Graphical_lasso_model}) when implementing gLasso such that the obtained sparsity level of the estimated inverse covariance matrix is approximately equal to the true value. We remark that both gLasso and PLT$_0$ only constrain the number of non-zero elements of the inverse covariance matrix to select significant conditional dependence between variables, and the graph constraints that arise from HEAs, such as node-type dependent interactions, are not considered.

Table \ref{RMSE_beta} reports the MSEs for all methods. The average number of non-zero elements of the true inverse covariance matrix $|\overline{\Sigma}^{-1}|$ is approximately $10p$. We can see that MSEs of CARGO decreases with a larger $|\overline{\Sigma}^{-1}|$ (or $p$) for most $\beta$ as well as the aggregated summary in the last column, while gLasso and PTL$_0$ tend to become similar for $p = 20$ to $p = 100$. CARGO outperforms the other two methods on nearly all individual $\beta$, with only a few exceptions. The last column of Table \ref{RMSE_beta} which averages over $\{\beta_1, \ldots, \beta_6\}$ indicates that CARGO gives the smallest MSEs on average for all $p$.
CARGO continues to improve with $p$ increased from 20 to 100, while gLasso and PTL$_0$ seem to stablize when $p$ attains 20 to 100, suggesting performance gain of CARGO by incorporating graph constraints. 
 \begin{table}[!htbp]
        \centering
        \caption{Comparison of three methods using MSE. Standard errors (SEs) across 50 simulations are reported. The last column reports the mean value of MSE and standard error averaged over $\beta_{1},...,\beta_{6}$. The smallest MSE among the three methods is marked in bold. All reported numbers are multiplied by 100. }
        \begin{tabular}{lccccccccc}
\toprule
$p$ ($|\overline{\Sigma}^{-1}|$)&method&  & $\beta_1$ & $\beta_2$ & $\beta_3$ &  $\beta_4$ & $\beta_5$ &  $\beta_6$ & aver.\\ 
\cmidrule{1-1} \cmidrule{2-10}
20& CARGO& MSE&0.49&0.89&\textbf{1.07}&0.84&0.46&\textbf{0.87}&\textbf{0.77}\\
(210)& &SE&0.12&0.14&0.13&0.12&0.09&0.10&0.12
\\  
 & gLasso& MSE&0.34&\textbf{0.78}&1.52&\textbf{0.74}&\textbf{0.29}&1.00&0.78
 \\
& & SE&0.10&0.13&0.21&0.12&0.07&0.12&0.11
\\
& PTL$_0$ &MSE&\textbf{0.17}&0.80&1.13&0.82&0.32&1.41&0.78 \\
& & SE&0.03&0.11&0.08&0.08&0.04&0.20&0.09
\\
\cmidrule{1-10}
50& CARGO & MSE&0.36&\textbf{0.68}&\textbf{0.93}&\textbf{0.62}&\textbf{0.21}&\textbf{0.98}&\textbf{0.63}\\
(495)& &SE&0.06&0.10&0.09&0.07&0.03&0.09&0.08
\\  
 &  gLasso & MSE&\textbf{0.21}&0.80&1.09&0.88&0.33&1.35&0.78
 \\
& & SE&0.02&0.05&0.06&0.04&0.02&0.04&0.04\\
& PTL$_0$ &MSE&0.23&0.82&1.13&0.94&0.33&1.40&0.81 \\
& & SE&0.02&0.05&0.08&0.03&0.02&0.04&0.04
\\
\cmidrule{1-10}
100& CARGO & MSE&0.23&\textbf{0.46}&\textbf{0.81}&\textbf{0.53}&\textbf{0.22}&\textbf{0.80}&\textbf{0.51}\\
(1039)& &SE&0.04&0.06&0.10&0.06&0.03&0.07&0.06
\\
 &  gLasso & MSE&\textbf{0.19}&0.85&1.04&0.97&0.34&1.43&0.80
 \\
& & SE&0.01&0.03&0.03&0.03&0.01&0.03&0.02\\
& PTL$_0$&MSE & 0.21&0.84&1.05&0.91&0.34&1.43&0.80\\
& & SE&0.02&0.04&0.03&0.03&0.02&0.03&0.03
\\
           \bottomrule
           \end{tabular}
           \vspace{0.05in}
        \label{RMSE_beta}
    \end{table}    
    
\begin{figure}[!htp]
\centering
\includegraphics[width = 0.45\linewidth]{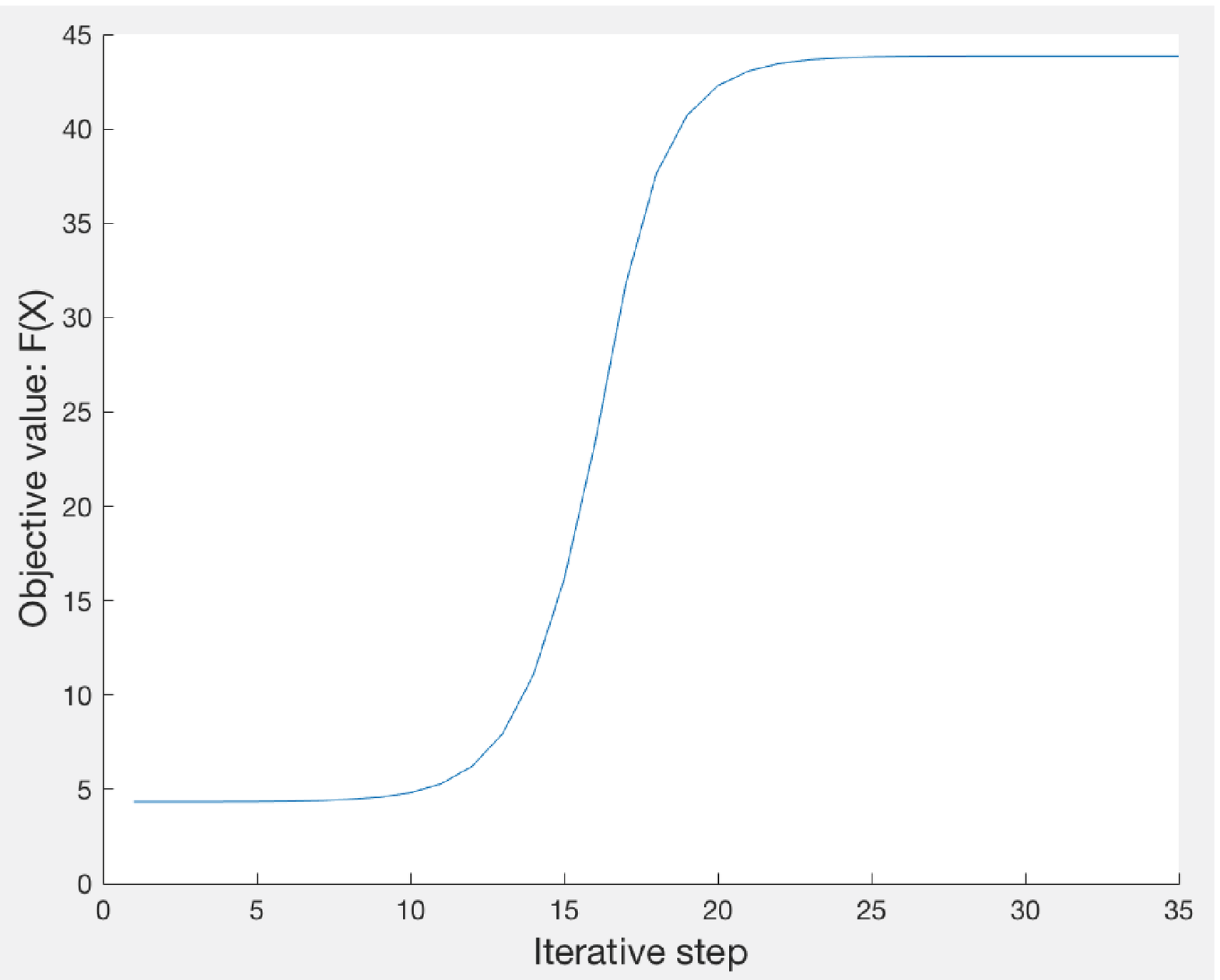} \hspace{20pt}
\includegraphics[width = 0.45\linewidth]{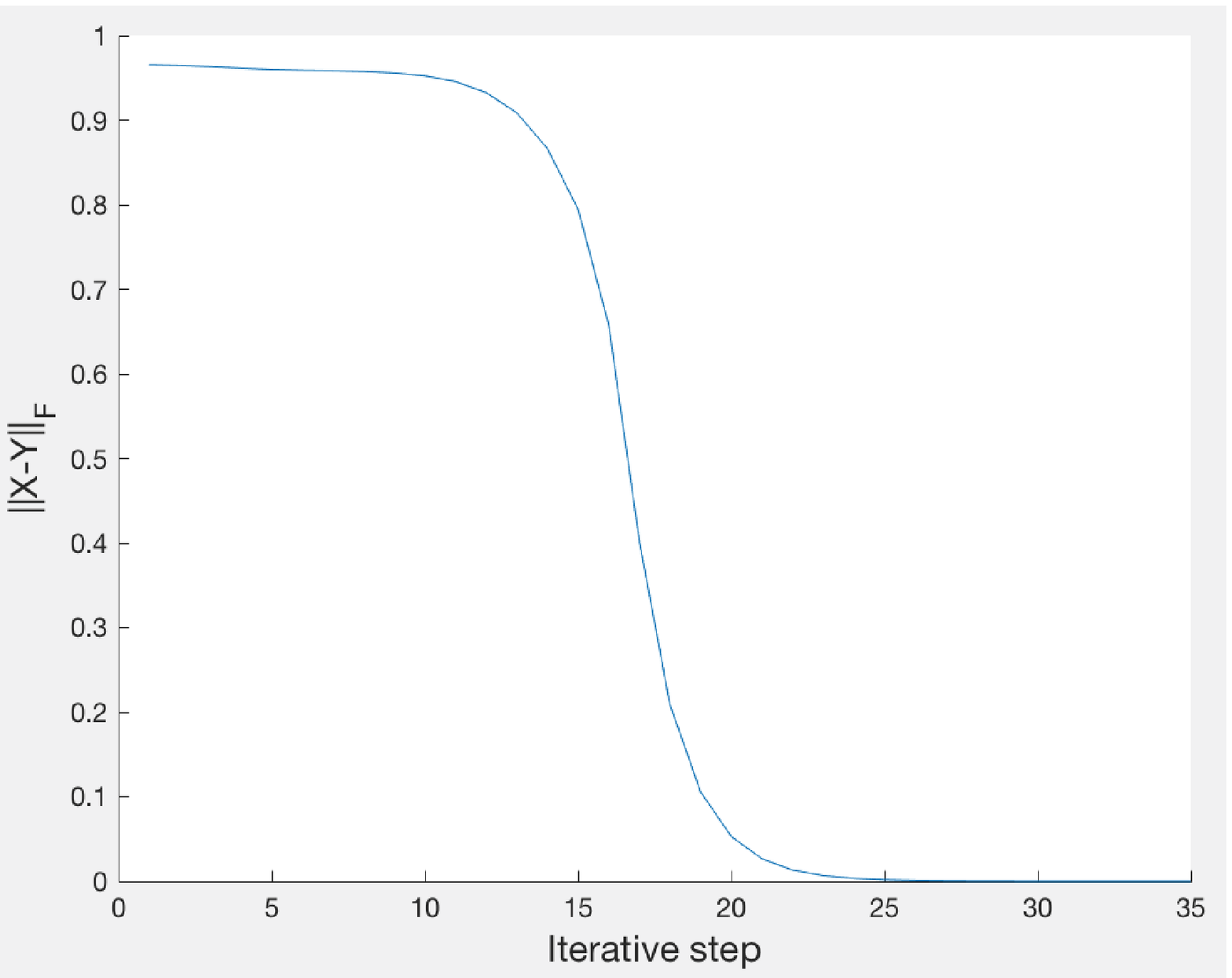}
\caption{Objective function $F(X)$ vs. iterative step (left) and the Frobenius norm of $X-Y$ vs. iterative step (right) when $p = 100$.}\label{converge_synthetic}
\end{figure}

We observe that the proposed optimization algorithm converges very fast. Choosing $p = 100$ as an example, the convergence diagnosis plots in Figure~\ref{converge_synthetic} show that both the objective function $F(X)$ and the Frobenius norm of $X-Y$ reach convergence in about 25 iterations, confirming the convergence analysis carried out in the preceding section. Convergence plots for $p = 20$ and $p = 50$ are similar and omitted here. 

\section{Application to HEAs} 
\label{sec:data.application}

In this section, we apply CARGO to analyze the novel HEA data described in Section~\ref{sec:background.HEA}. The data consist of three different HEA structures (fcc, hcp, and dhcp) and eight samples for each structure. We analyze these structures and samples individually to study the variability and transferability across samples and structures.

We apply the proposed method to detect the magnetic interactions between Ni, Co, Cr, Fe atoms. This gives estimates of a 10-dimensional parameter $\bm{\beta} = (\beta_1, \ldots, \beta_{10})$, corresponding to the interaction in atom pairs Ni-Ni, Co-Co, Cr-Cr, Fe-Fe, Ni-Co, Ni-Cr, Ni-Fe, Co-Cr, Co-Fe, Cr-Fe, respectively. 

Table \ref{fcc_estimates}-\ref{dhcp_estimates} report the results from eight samples for the fcc, hcp and dhcp structure.   We can see that $\bm{\beta}$ estimated from individual samples with radically different atom type orders tend to show the same signs. In particular, the interaction in the Ni-Ni, Co-Co, Fe-Fe, Ni-Co, Ni-Fe, Co-Fe pairs are positive in most cases, that is, there is a positive contribution for magnetic moment between the atoms Ni, Co and Fe. In contrast, the interactions in Cr-Cr, Ni-Cr, Co-Cr, Cr-Fe pairs are negative in most cases, suggesting the presence of atom Cr tends to decrease the magnetic moment of atom Ni, Co and Fe in the interacting pairs. Thus, our analysis shows that the atom interaction is always negative once containing atom Cr, which implies that atom Cr will result in the reduction of the magnetic moments of the overall high-entropy alloy. This observation has been confirmed in the existing literature as the magnetic frustration effect of Cr~\citep{Huang2016Mechanism, Schneeweiss2017Magnetic, Niu2018Magnetically}, where Ni-Cr, Co-Cr and Fe-Cr pairs reflect an anti-ferromagnetic coupling due to the fact that the Cr magnetic moment is anti-parallel to that of Ni, Co and Fe. CARGO is model-based and favorably parsimonious, and such alignment between CARGO and the literature solely based on domain knowledge and intricate physical laws is reassuring. 

The last column averages across 8 samples for each HEA structure. We observe that the largest three interactions in absolute value are Fe-Fe, Co-Fe, and Cr-Fe for the three structures, suggesting profound interactions between Fe the other elements. While this pattern persists for three structures, the quantification of such interactions vary from sample to sample, and from structure to structure. This model-based quantification by CARGO substantially expands qualitative descriptions that are available in the literature. Furthermore, the proposed method is favorably parsimonious, leading to improved interpretability. Table \ref{fcc_estimates}-\ref{dhcp_estimates} also show the variability of the estimated interactions across samples, although they tend to be similar and are at least at the same magnitude. This observation substantiates the use of single-sample models when the number of samples is extremely limited compared to the enormous space of all arrangements of atom types. 
\begin{table}[!htbp]
    \centering
	\caption{Estimates of $\bm{\beta}$ from 8 independent samples for HEAs with fcc structure.} \label{fcc_estimates}
		\setlength{\tabcolsep}{0.8mm}{
		\begin{tabular}{@{}crrrrrrrrr@{}}
		\toprule
\multirow{2}{*}{Parameter} & \multicolumn{8}{c}{Sample ID} & \multirow{2}{*}{Mean} \\
\cmidrule{2-9} 
  & 1  & 2 & 3 & 4 & 5 & 6 & 7 & 8 &  \\
\cmidrule{1-1} \cmidrule{2-9} \cmidrule{10-10}   
Ni-Ni&0.0359&0.0319&0.0210&0.0172&0.0126&0.0272&-0.0081&0.0350&0.0216\\
Co-Co&0.2900&0.1337&-0.0241&0.0965&0.2988&0.0329&-0.0889&0.0684&0.1009\\
Cr-Cr&-0.1085&0.0024&0.0929&-0.1106&-0.0054&-0.1028&0.0644&0.0979&-0.0087\\
Fe-Fe&0.5183&0.6405&0.5475&0.4672&0.6436&0.4979&0.4581&0.4433&0.5270\\
Ni-Co&0.1077&0.1345&0.0675&0.0692&0.0971&0.1209&0.0246&0.1209&0.0928\\
Ni-Cr&-0.0364&-0.0253&-0.0801&-0.0561&-0.0293&-0.0258&0.0232&0.0404&-0.0237\\
Ni-Fe&0.0567&0.0529&0.0250&0.0764&0.0762&0.0527&0.1277&0.0384&0.0632\\
Co-Cr&-0.0796&-0.0833&0.0678&-0.1692&-0.0560&-0.1021&-0.1171&-0.1446&-0.0855\\
Co-Fe&0.2195&0.2266&0.3148&0.3390&0.2513&0.2793&0.3851&0.2426&0.2823\\
Fe-Fe&-0.2537&-0.1756&-0.1943&-0.1044&-0.1424&-0.2432&-0.1016&-0.1007&-0.1645\\
          \bottomrule
          \end{tabular}}
\newline
\vspace*{0.3 cm}
	\caption{Estimates of $\bm{\beta}$ from 8 independent samples for HEAs with hcp structure.} \label{hcp_estimates}
		\setlength{\tabcolsep}{0.8mm}{
		\begin{tabular}{@{}crrrrrrrrr@{}}
		\toprule
\multirow{2}{*}{Parameter} & \multicolumn{8}{c}{Sample ID} & \multirow{2}{*}{Mean} \\
\cmidrule{2-9} 
  & 1  & 2 & 3 & 4 & 5 & 6 & 7 & 8 &  \\
\cmidrule{1-1} \cmidrule{2-9} \cmidrule{10-10}  
Ni-Ni&0.0059&-0.0253&0.1160&0.0234&0.0334&0.0075&0.0012&0.0545&0.0271\\
Co-Co&0.3723&0.1043&0.0902&0.0976&0.2217&0.2098&	0.3661&0.1547&0.2021\\
Cr-Cr&-0.0675&0.0422&-0.0224&-0.0882&-0.0198&0.0105&-0.1559&0.0868&-0.0268\\
Fe-Fe&0.0143&0.1372&0.1695&0.5059&0.6039&0.0786&0.1688&0.5107&0.2736\\
Ni-Co&0.0576&0.0071&0.0732&0.1762&0.0604&0.0453&0.1078&0.1010&0.0786\\
Ni-Cr&0.0026&-0.0278&-0.0554&0.0400&-0.0585&-0.0618&0.0125&-0.0318&-0.0225\\
Ni-Fe&0.1225&0.1329&0.0608&0.0414&0.0788&0.0910&-0.0148&0.0865&0.0749\\
Co-Cr&-0.0974&-0.1323&-0.0492&-0.0202&-0.1576&-0.0218&-0.0745&-0.0638&-0.0771\\
Co-Fe&0.2103&0.3017&0.2077&0.2720&0.2365&0.3330&0.0317&0.2821&0.2344\\
Cr-Fe&-0.2269&-0.0823&-0.1254&-0.2039&-0.1787&-0.1011&-0.1373&-0.1903&-0.1557\\
          \bottomrule
          \end{tabular}} 
\newline
\vspace*{0.3 cm}

	\caption{Estimates of $\bm{\beta}$ from 8 independent samples for HEAs with dhcp structure.} \label{dhcp_estimates}
		\setlength{\tabcolsep}{0.8mm}{
		\begin{tabular}{@{}crrrrrrrrr@{}}
		\toprule
\multirow{2}{*}{Parameter} & \multicolumn{8}{c}{Sample ID} & \multirow{2}{*}{Mean} \\
\cmidrule{2-9} 
  & 1  & 2 & 3 & 4 & 5 & 6 & 7 & 8 &  \\
\cmidrule{1-1} \cmidrule{2-9} \cmidrule{10-10}
Ni-Ni&-0.0023&0.0154&-0.0067&0.0282&0.0048&-0.0108&-0.013&0.0024&0.0023\\
Co-Co&0.3851&0.1811&0.2268&0.2927&0.1587&0.2758&0.1927&0.1691&0.2352\\
Cr-Cr&-0.0796&0.0216&-0.0700&-0.0318&-0.0332&0.0192&-0.0490&-0.0403&-0.0329\\
Fe-Fe&0.4386&0.5830&0.4947&0.5857&0.5723&0.4591&0.4949&0.1200&0.4685\\
Ni-Co&0.0807&0.0264&0.0177&0.0785&0.0374&0.0125&0.0081&0.0435&0.0381\\
Ni-Cr&-0.0067&-0.0208&-0.0299&0.0103&-0.0061&-0.0052&-0.0239&0.0041&-0.0098\\
Ni-Fe&0.0810&0.1068&0.1155&0.0850&0.0914&0.1350&0.1056&0.1057&0.1032\\
Co-Cr&-0.0896&-0.0813&-0.0772&-0.0340&-0.0686&-0.0294&-0.0893&-0.0069&-0.0595\\
Cr-Fe&0.1568&0.2653&0.2537&0.2498&0.2420&0.2628&0.2470&0.2297&0.2384\\
Fe-Fe&-0.1587&-0.1578&-0.1220&-0.1762&-0.1580&-0.1100&-0.1501&-0.1261&-0.1449\\
          \bottomrule
          \label{table6}
          \end{tabular}}
\end{table}

\begin{figure}[!htbp]
\centering
\includegraphics[width=0.45\linewidth]{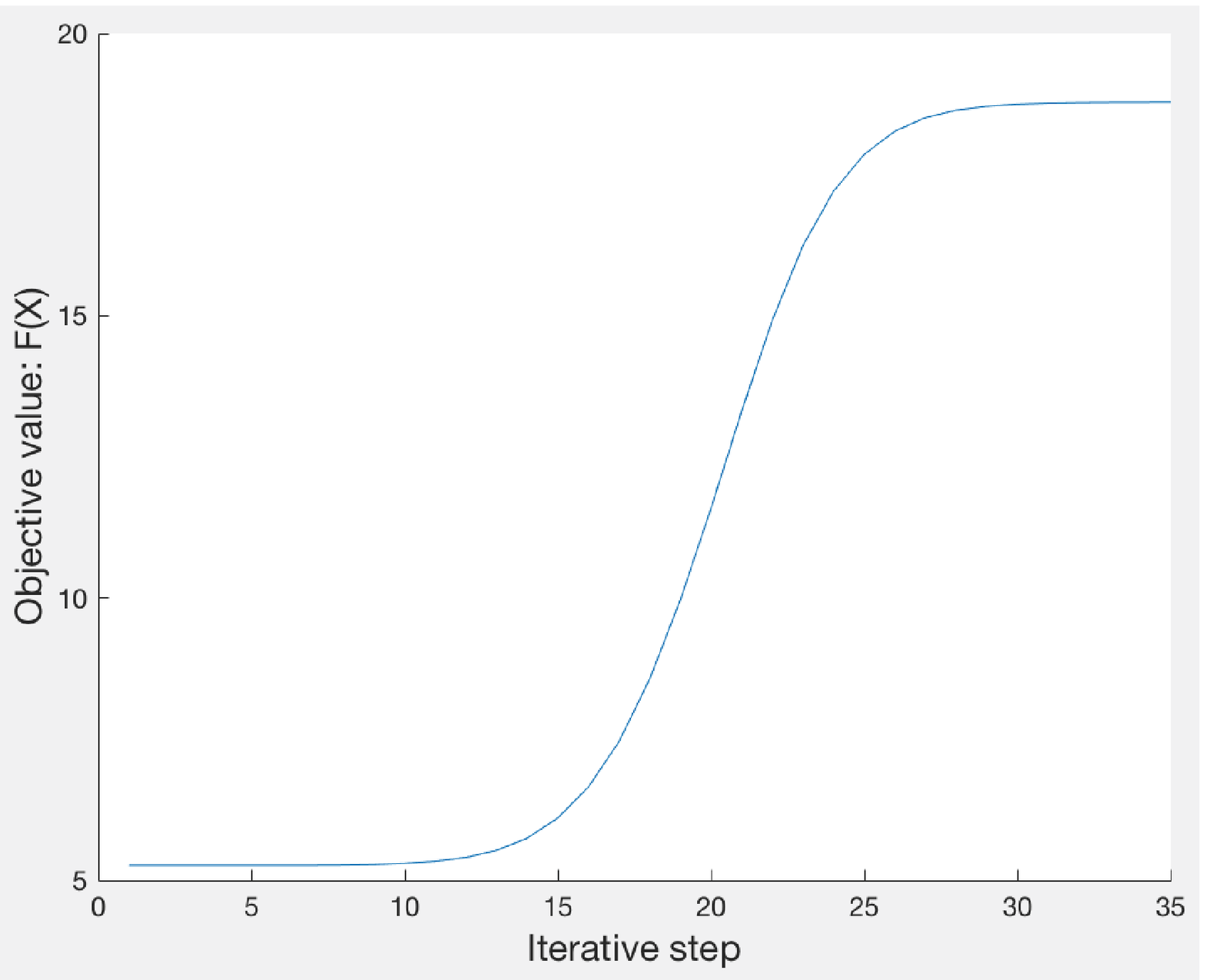} \hspace{20pt}
\includegraphics[width=0.45\linewidth]{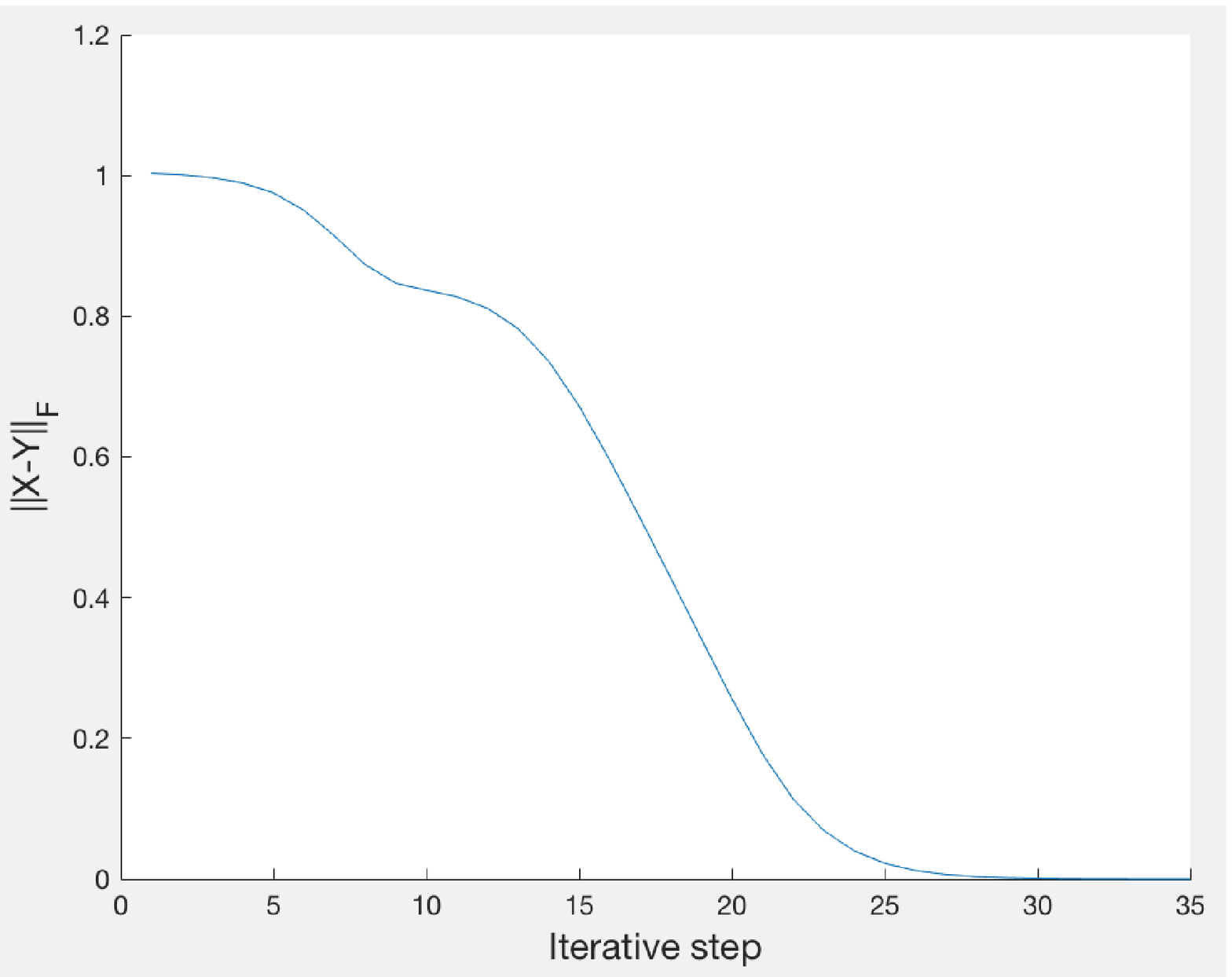}
\caption{Objective function $F(X)$ vs. iterative step (left) and the Frobenius norm of $X-Y$ vs. itertive step (right) for one sample.}\label{fig:convergence.real.data}
\end{figure}

Figure~\ref{fig:convergence.real.data} shows the convergence diagnosis plots, in which both the objective function $F(X)$ and the Frobenius norm of $X-Y$ converge fast in about 25 iterations. This is similar to the observation made in the simulation, and confirms our convergence analysis in Section~\ref{sec:algorithm.convergence}. 

\section{Discussion}
\label{sec:discuss}

In this article, we study inverse covariance matrix estimation with graph constraints. We introduce novel formulation to encode structural constraints encompassing node types, topological distance of node, and partially specified conditional dependence patterns in graphical models. This leads to a Bayesian conditional autoregressive model with graph constraints (CARGO) to study conditional dependence of data via graphical models with intricate constraints that arise from domain knowledge. We develop efficient implementation of CARGO through a modified Gauss-Seidel scheme for posterior exploration, and provide closed-form estimates using promixity operators. A range of algorithmic convergence results for the proposed algorithm are provided, enabling fast computation. 

In a novel real data application to HEAs, the proposed methods lead to model-based quantification and interpretation of magnetic moment interactions with high tractability and transferability, providing data-driven insights to the \emph{magnetism}-property-processing relationships in HEA designs and complementing first principles simulations. To our best knowledge, CARGO is the first model-based approach to quantify magnetic moment interactions in HEAs. There are several interesting next directions building on our initial CARGO approach. Firstly, it might be of practical relevance to extend the proposed method to multi-sample and multi-structure models that incorporate
sample-level and structure-level random effects via hierarchical modeling. It is also interesting to develop regression models to relate the structured inverse covariance matrices to other important properties of HEAs, such as formation energy, elastic moduli, and stacking-fault energies.

\bibliography{HEAs}

\end{document}